\documentclass[preprint,12pt]{elsarticle}

\usepackage{amssymb}
\usepackage{amsmath}
\usepackage{graphicx}
\usepackage{epsfig,latexsym}
\usepackage{slashed}
\usepackage{geometry}
\usepackage{simplewick}
\usepackage{caption}
\usepackage{subcaption}
\usepackage{nicefrac}
\usepackage[colorlinks=true,linktocpage=true,linkcolor=blue,citecolor=blue]{hyperref}
\usepackage[usenames,dvipsnames]{color}
\usepackage[normalem]{ulem}

\def\be{\begin{eqnarray*}}
\def\ee{\end{eqnarray*}}
\def\beq{\begin{eqnarray}}
\def\eeq{\end{eqnarray}}
\def\bem{\begin{multline}}
\def\eem{\end{multline}}

\def\vac{\text{vac}}
\def\med{\text{med}}
\def\gain{\text{gain}}
\def\loss{\text{loss}}
\def\coh{\text{coh}}

\def\tot{\text{tot}}
\def\cut{\text{cut}}

\def\jet{\text{jet}}
\def\dd{\text{d}}
\def\abar{\bar\alpha}
\def\QW{\text{\tiny QW}}
\def\BDMPS{\text{\tiny BDMPS}}
\def\thetajet{R}
\def\abar{\bar\alpha}

\newcommand{\onehalf}{{\nicefrac{1}{2}}}
\newcommand{\threehalfs}{{\nicefrac{3}{2}}}

\newcommand{\pT}{p_\text{\tiny T}}

\newcommand{\pTone}{p_{\text{\tiny T} 1}}
\newcommand{\pTtwo}{p_{\text{\tiny T} 2}}
\newcommand{\rme}{\text{e}}
\newcommand{\rmd}{\text{d}}
\newcommand{\pTg}{p_\text{{\tiny T}g}}

\def\nn{\nonumber \\}

\def\zcut{z_\text{cut}}

\journal{Physics Letters B}

\begin{document}
\begin{frontmatter}

\title{Groomed jets in heavy-ion collisions: sensitivity to medium-induced bremsstrahlung}

\author[a]{Yacine Mehtar-Tani}
\ead{ymehtar@uw.edu}

\author[b]{Konrad Tywoniuk}
\ead{konrad.tywoniuk@cern.ch}

\address[a]{Institute of Nuclear Theory, University of Washington, Seattle, WA 98195-1550, USA}
\address[b]{Theoretical Physics Department, CERN, 1211 Geneva 23, Switzerland}

\begin{abstract}
We argue that contemporary jet substructure techniques might facilitate a more direct measurement of hard medium-induced gluon bremsstrahlung in heavy-ion collisions, and focus specifically on the ``soft drop declustering''  procedure that singles out the two leading jet substructures.
Assuming coherent jet energy loss, we find an enhancement of the distribution of the energy fractions shared by the two substructures at small subjet energy caused by hard medium-induced gluon radiation. Departures from this approximation are discussed, in particular, the effects of colour decoherence and the contamination of the grooming procedure by soft background. Finally, we propose a complementary observable, that is the ratio of the two-pronged probability in Pb-Pb to proton-proton collisions and discuss its sensitivity to various energy loss mechanisms.

\end{abstract}
\begin{keyword}
QCD jets \sep jet quenching \sep jet grooming
\end{keyword}
\end{frontmatter}

\begin{flushright}
INT-PUB-16-038, CERN-TH-2016-227
\end{flushright}
%

\section{Introduction}
\label{sec:intro}

The study of jets has become one of the core activities aiming to probe the properties of the quark-gluon plasma in heavy ion collisions at RHIC and, particularly in the last few years, at the LHC. While the experimental efforts initially focussed on measuring hadron spectra at large transverse momentum, the field has evolved into providing a wealth of measurements using reconstructed jets at the two colliding energies \cite{d'Enterria:2009am,Majumder:2010qh,Muller:2012zq}. In addition to measuring inclusive jet spectra \cite{Chatrchyan:2012nia,Aad:2014bxa,Abelev:2013kqa}, which are sensitive the magnitude of energy loss, detailed measurements of jet substructure \cite{Chatrchyan:2012gw,Chatrchyan:2013kwa} and large-angle energy flow shed light on the intricate nature of the jet interactions with the dense, deconfined QCD matter formed in the collisions.

These modifications are expected to arise from a complex interplay of elastic and inelastic processes which alter the distribution within the jet cone and, ultimately, propagate a fraction of the total jet energy to large angles. However, lacking a rigorous theoretical framework for jet evolution in matter, current, state-of-the-art models rely on various assumptions when interfacing medium effects with jet showering algorithms. The convolutedness of such a task allows generic models to describe qualitative features of the data, calling into question the discriminative power of such observables. In order to achieve a consistent understanding of the underlying mechanisms responsible for the observed modifications, several questions remain to be answered. Do the jet constituents lose energy coherently or as independent colour charges? Is energy loss predominantly perturbative, and hence caused by medium-induced radiation, or dominated by direct energy injection into the thermal bath via non-perturbative drag? Finally, to what extent does the medium response to the jet propagation and fragmentation affect jet observables? 

These issues prompt us to explore new observables which could shed more light on the details of the jet evolution. We argue below that a set of jet substructure observables that pin down details of the ``first'' hard splitting, which will be defined shortly, provide a possibly more direct measurement of hard medium-induced radiation. This could provide a first observation of the generic mechanism that is expected to drive jet modifications, in general, and energy loss, in particular.\footnote{Energy loss can also be caused entirely by elastic collisions (drag).} 

Jet grooming techniques (such as trimming, pruning etc., see \cite{Dasgupta:2013via,Dasgupta:2013ihk}) have recently been developed and are extensively studied at the LHC. They provide useful tools for further quantifying QCD jet substructure, for reviews see \cite{Salam:2009jx,Altheimer:2012mn}. These techniques are generally designed to single out perturbative radiation from soft, mostly non-perturbative components of the jet. In this work, we  focus on the ``soft drop declustering'' procedure \cite{Larkoski:2014wba} which consists of sequentially declustering the jet constituents using the Cambridge-Aachen algorithm (C/A) back to the first, hard splitting. In effect, the procedure terminates when the first pair of subjets that satisfy the grooming condition,
\beq
\label{eq:CutCondition}
\frac{\min(\pTone, \pTtwo)}{\pTone+\pTtwo} > \zcut \left(\frac{\Delta R_{12}}{R} \right)^\beta \,,
\eeq
is identified, where $\pTone$ ($\pTtwo$) are the two subjet transverse momenta and $\Delta R_{12}$ their angular separation ($R$ being the jet radius). Hence, the declustering procedure {\it de facto} picks out two substructures that pass the condition at the largest angular separation.
The kinematics of that pair defines the jet groomed momentum $\pTg \equiv \pTone + \pTtwo $, which is smaller that the original jet $\pT$, the sharing variable $z_g \equiv \min(p_{\text{\tiny T} 1},p_{\text{\tiny T} 2})/\pTg$ and groomed jet radius $r_g \equiv \Delta R_{12}/R$ \cite{Larkoski:2014wba}. The normalized $z_g$-distribution is referred to as the ``splitting probability'' and has been studied recently \cite{Larkoski:2015lea}. At leading-log accuracy, jet evolution in vacuum is ordered in angle and, therefore, also in formation time. This implies that subsequent emissions during the course of the jet evolution take place at diminishing angles. Hence, the soft drop identifies the first pair in the jet evolution history. 

This observable has, for the first time, been successfully measured at the LHC for $\beta = 0$ with the implementation of a minimal resolution angle $\Delta R_{12} \geq 0.1$ to account for detector resolution effects \cite{CMS:2016jys}. The $z_g$-distribution appears to be steeper in heavy-ion collisions than in proton-proton collisions and the effect seems to decrease with increasing jet energy. 

In this work, we analyse the measured $z_g$-distribution and propose a new and complementary observable: the probability of finding one or two-pronged structures given a resolution angle $R_0$, that we argue is sensitive to the details of jet-medium interaction. We show in particular that in-cone medium-induced radiation marginally increases the the two-pronged probability,  while   the effect of incoherent energy loss yields a strong suppression. 

We structure the paper in the following way. First, in Sec.~\ref{sec:vacuum}, we formally introduce the observables of interest, that is, the $z_g$-distribution of the primary two subjets, in the Modified-Leading-Log approximation (MLLA) \cite{Larkoski:2014wba}, and the integrated  probability to measure one- or two-pronged structures given a resolution angle $R_0$.  In Sec.~\ref{sec:Eloss-coherence}, we leave aside the collinear vacuum radiation and discuss multiple medium-induced gluon bremsstrahlung that highlights the relevance of two typical regimes: rare, in-cone hard gluon radiation which may be directly probed by the grooming procedure and an out-of-cone soft gluon cascade that causes the jet to lose energy. We argue that the range of variables explored by the CMS collaboration justifies this angular separation. 
Section~\ref{sec:Observable} contains our main results, namely the effects of energy loss and hard radiation for coherent jets. In this scenario, we resum the effect of multiple collinear vacuum splittings.
Finally, we discuss alternative scenarios that leaves out either of these two main components in Sec.~\ref{sec:alternative}, i.e., incoherent energy loss and the effect of medium back-reaction on the observable.  This allows us to draw generic conclusions about how these mechanisms are expected to affect the observable.  Contrasting the coherent and incoherent scenarios disclose for the first time the strong medium effects on resolved jet structures. We summarise and close by presenting an outlook in Sec.~\ref{sec:outlook}.

\section{The splitting function in vacuum }
\label{sec:vacuum}

The splitting probability of the longitudinal energy fraction $z_g$ of a groomed jet in the vacuum \cite{Larkoski:2015lea}, see also \cite{Larkoski:2014wba}, reads
\beq
\label{eq:SplitFuncDef}
\textsl{p}(z_g) = \int_0^{\thetajet} \dd \theta \, \Delta(\thetajet, \theta) \mathcal{P}_\vac(z_g,\theta)\, \Theta\left(z-z_\cut \theta^\beta \right)\,,
\eeq
for $z_g>1/2$, where $\thetajet$ is the opening angle of the jet and the Heaviside step function embodies the condition in Eq.~(\ref{eq:CutCondition}). 
At leading order, the splitting function is given by
\beq
\label{eq:VacuumSplitting}
\mathcal{P}_\vac (z,\theta) = \bar \alpha \frac{P(z)}{\theta}  \,,
\eeq
where $P(z)$  is the relevant Altarelli-Parisi splitting function. For simplicity and without loss of generality, we shall restrict our discussion to the purely gluonic case, i.e., $P(z)\equiv P_{gg}(z) = (1-z(1-z))^2/(z(1-z))$ with $\bar\alpha= 2  \alpha_s N_c/\pi$.

By itself, Eq.~(\ref{eq:VacuumSplitting}) describes a single gluon splitting and is not a satisfactory definition of a physical observable since it contains both soft ($z\to 0$) and collinear ($\theta \to 0$) divergences. In order to establish such a definition, we have to include multiple gluon emissions in the Sudakov form factor that corresponds to the probability not to measure a splitting that would fulfil the soft drop condition (\ref{eq:CutCondition}), between $R$ and $\theta$,
\beq\label{eq:Sudakov}
\Delta(\thetajet,\theta) = \exp \left[-\abar \int_\theta^{\thetajet}\frac{\dd \theta' }{\theta'}\int_0^{1/2} \dd z \,P(z) \Theta(z-z_\cut \theta'^\beta) \right].
\eeq
For further details, see \ref{sec:AppendixSudakov}.
It is a straightforward exercise to show, given the definitions above, that $\int_0^{\onehalf} \dd z_g\, \textsl{p}(z_g)=1$. Hence,  Eq.~(\ref{eq:VacuumSplitting}) is indeed a probability.
For different values of $\beta \geq 0$ the soft drop removes various degrees of soft and soft-collinear radiation. For $\beta <0$, even purely collinear radiation in the jet is removed.\footnote{Throughout the paper, we present formulas for arbitrary $\beta$, but only consider $\beta = 0$ in order to compare the obtained features with experimental data.}

If now we require a minimal resolution angle,  $\theta > R_0$, it implies that a number of splittings will be detected as single prong objects and, hence, must be rejected. As a result, $\textsl{p}(z_g)$ is now normalised to the probability of resolving two-pronged structures,
\beq
\label{eq:2prongVacuum}
\mathbb{P}_{2\text{prong}} = \int_0^{1/2} \dd z_g\, \textsl{p}(z_g) =  1- \Delta(R,R_0) \,,
\eeq
where we have replaced the lower limit in Eq.~(\ref{eq:SplitFuncDef}) by $R_0$. In other words, the probability of observing a groomed jet with a single constituent given a resolution angle $R_0$ is simply $\mathbb{P}_{1\text{prong}}=\Delta(R,R_0)$, so that $\mathbb{P}_{1\text{prong}}+\mathbb{P}_{2\text{prong}}=1$. These considerations are quite general and do not depend on the specific features of the grooming procedure.

For the computation of the corresponding splitting probability (\ref{eq:SplitFuncDef}) of jets produced in heavy-ion collisions one has to take into consideration several important aspects. First, one has to examine the effects of  energy loss on the jet. This is, e.g., reflected in the suppression of the inclusive jet yield and is well understood as arising mainly from large-angle multiple soft emissions. This turns generally out to be a complicated task, since the probability explicitly relies on multiple emissions through the grooming procedure. Ultimately, both the splitting process and the corresponding Sudakov form factor could be affected. In order to settle the issue, one has to consider the number of jet structures that are resolved by the medium. In this simplest case, the jet is unresolved and therefore interacts coherently with the medium. We focus on this case in Sec.~\ref{sec:Observable}, leaving a discussion of more involved situations for Sec.~\ref{sec:alternative}.

\section{Radiative energy loss and colour coherence}
\label{sec:Eloss-coherence}

Before we proceed with the analysis of the observable of interest, let us first revisit the inelastic processes that we consider in the present work. 

\subsection{Medium-induced rare in-cone radiation vs out-of-cone soft cascade}
\label{sec:MediumInduced}

It was early realised that energetic partons traversing a coloured medium would lose energy through inelastic interactions with the medium, or so-called induced bremsstrahlung. 
For further details and refinements, we refer the interested reader to Refs.~\cite{Mehtar-Tani:2013pia,Blaizot:2015lma}.
The BDMPS-Z spectrum of hard primary emitted gluons \cite{Baier:1996kr,Baier:1996sk,Zakharov:1996fv,Zakharov:1997uu} (see also \cite{Blaizot:2012fh} for a recent discussion) reads 
\beq
\label{eq:BDMPSspectrum}
 \frac{\rmd N_\BDMPS}{\rmd z} = \bar \alpha   P_{gg}(z) \ln\left|\cos \frac{1+i}{2} \sqrt{\frac{\omega_c}{\pT}\frac{1+z(1-z)}{z(1-z)}} \right| \,,
\eeq
where $z\equiv\omega/\pT$ is the jet energy fraction carried by the daughter gluon. Due to the so-called Landau-Pomeranchuk-Migdal (LPM) interference, the energy spectrum of induced gluons scale is suppressed below the characteristic energy $\omega_c \equiv \hat q L^2/2$, as $\omega \dd N_\BDMPS/\dd \omega \simeq \bar
\alpha\sqrt{\omega_c/ \omega}$, where $\hat q$ is the medium transport parameter and $L$ its length.\footnote{Here it is worth commenting, that even if we discuss an ``enhancement'' of the splitting probability, the medium-induced LPM spectrum is always suppressed compared to the expected spectrum arising from independent inelastic scattering in the medium. The possibility of enhancement arises due to the stronger (apparent) infrared (IR) divergence of the LPM spectrum and the relative weights between vacuum and medium-induced emissions that can, in effect, modify the functional behaviour of the splitting probability.} When $\omega>\omega_c$, the suppression is even stronger, i.e., $1/\omega^2$. 
This cut-off scale also determines the average energy of emitted gluons. Furthermore, the BDMPS-Z spectrum must be cut-off in the infrared at the Bethe-Heitler frequency $\omega_\text{BH} \sim \hat q \ell^2_\text{mfp}$, which signals the regime where the gluon formation time becomes comparable to the elastic mean free path $\ell_\text{mfp}$. Hence, for this picture to be valid we must have $\ell_\text{mfp}\ll L$. 

The multiplicity of emitted gluons is however governed by gluons with energy $\omega \lesssim \omega_s$, where $\omega_s \equiv \bar\alpha^2 \omega_c$ \cite{Baier:2001yt}. They are emitted copiously, $\int_{\omega_s}\dd \omega\, \dd N_\BDMPS/\dd \omega \gtrsim \mathcal{O}(1)$, and have to be resummed. Furthermore, since their formation time is much shorter than the medium length, secondary branchings have to be taken into account in order to trace the final distribution of energy \cite{Blaizot:2013hx,Blaizot:2013vha}. This distribution is evolved in time, in contrast to the resummation of vacuum emission in MLLA, and its angular structure has also been studied in detail \cite{Blaizot:2014ula,Blaizot:2014rla,Kurkela:2014tla}.

It is instructive to estimate the relevant scales for typical values of the parameters. For $L=4$ fm, $\hat q\simeq 1$ GeV$^2$/fm and $\bar \alpha \simeq 0.3$, we find $\omega_c$ = 80 GeV and  $\omega_s \simeq 7$ GeV.  Hence, frequencies $\omega>\omega_s$ correspond to rare emissions whose probability is given by the leading order BDMPS-Z spectrum. Furthermore, their characteristic emission angle is $\theta_\BDMPS =k_\perp/\omega \sim \sqrt{\hat q L}/\omega$, which for 7 GeV  $< \omega < $ 80 GeV yields $0.025 < \theta_\BDMPS < 0.28$. It follows that for this set of parameters the hard BDMPS-Z gluons are radiated within the jet cone $R= 0.3$.   Thus medium-induced gluons with long formation times, that can be radiated within the jet cone, could be identified as the ``first'' splitting, according to the soft drop because of their relatively large emission angle compared to that in vacuum, due to the LPM suppression (cf. the $
\theta$ factor in Eq.~(\ref{eq:BDMPSfinal})). Hence, they would contribute to a modification of the $z_g$-spectrum compared to the vacuum expectation.\footnote{This enhancement goes like $z^{-\threehalfs}$ in the multiple scattering regime and $z^{-2}$ for a thin medium \cite{Gyulassy:2000fs,Gyulassy:2000er}, see also \cite{Chien:2016led}. Interestingly enough, in both cases the main contribution to the enhancement arises in the LPM regime.} 

As mentioned above, when $\omega < \omega_s$, multiple branching are highly probable and as a result these soft gluons fragment rapidly in the medium ending up at relatively large angles $\theta_\text{soft} > \bar\alpha^{-2} (\hat q L^3)^{-1/2} \simeq 0.28$.   
This gluon cascade transports energy to large angles causing energy degradation \cite{Blaizot:2014ula,Blaizot:2014rla,Kurkela:2014tla}. It follows that a high energy parton loses energy at large angles by radiating multiple soft gluons. The probability to lose $\epsilon$ energy along a pathlength $L$, will be denoted by $D_\QW(\epsilon)$, where the subscript stand for quenching weights (QW) \cite{Baier:2001yt,Salgado:2003gb,Baier:2006fr}. In this work we will use the simple analytical form,
\beq
\label{eq:QW-prob}
D_\QW(\epsilon) = \sqrt{\frac{\omega_s}{\epsilon^{3}}} \, \rme^{- \frac{\pi \omega_s}{\epsilon}}. 
\eeq
It relates the jet spectrum in the presence of a medium to that in vacuum, $\dd N_\text{jet(0)}/ \dd p^2_{\tiny T}$, which is equal to the jet spectrum in proton-proton collisions scaled by the number of binary nucleon-nucleon collisions, 
\beq
\frac{\dd N_\text{jet}}{\dd \pT^2} = \int_{0}^\infty \dd \epsilon \, D_\QW\left(\epsilon \right) \frac{\dd N_\text{jet(0)}(\pT+\epsilon)}{\dd \pT^2}.
\eeq
The suppression of the jet spectrum is then given by the so-called quenching factor,
\beq
\label{eq:QuenchingFactor}
Q(\pT)= \int_{0}^\infty \dd \epsilon \, D_\QW\left(\epsilon\right) \frac{\dd N_\text{jet(0)}(\pT+\epsilon)}{\dd \pT^2} \Bigg/ \frac{\dd N_\text{jet(0)}}{\dd \pT^2} \,,
\eeq
measured experimentally by the nuclear modification factor, $R_{AA} \simeq Q(\pT)$. 
For a steeply falling jet spectrum, i.e. $\dd N_{\jet(0)}/\dd \pT^2 = \pT^{-n}$, 
with $n\gg 1$, we recover the well-known expression $Q(\pT) \simeq \exp[ - 2\sqrt{\pi n \omega_s/\pT} ]$ \cite{Baier:2001yt}, that shows that energy loss is dominated by multiple soft gluon radiation with energy $\omega_s$. 

\subsection{Colour coherence and energy loss}

While the above discussion concerns how energy is taken away from a single propagating colour constituent via radiative processes, recently it was realised how multi-parton configurations, relevant for jet formation, radiate in the medium \cite{MehtarTani:2010ma,MehtarTani:2011tz,MehtarTani:2011gf,MehtarTani:2012cy,CasalderreySolana:2011rz}. For a time-like colour dipole, with opening angle $\theta_0$, one can identify a critical angle $\theta_c \sim \sqrt{12/\hat q L^3}$ which separates two regimes. In the coherent regime $\theta_0 \ll \theta_c$, the medium only resolves the total colour charge of the dipole. In this case, energy is lost coherently by the system while subsequent fragmentation occurs as in the vacuum. In the opposite, incoherent case, the colour correlation of the constituents is broken. See also \cite{Casalderrey-Solana:2015bww,Arnold:2016kek} for similar conclusions in the context of the two-gluon emission spectrum.
These findings imply that energy loss should be applied only to substructures of the jet that are resolved by the medium \cite{CasalderreySolana:2012ef}. Taken at face value, it also suggests that resolved jets, i.e., jets consisting of several medium-resolved substructures, suffer more violent modifications than the ones that contain most of their energy within an unresolved core. This minimalistic approach was successfully applied to the understanding of jet quenching and modified fragmentation functions \cite{Mehtar-Tani:2014yea}.

We adopt a similar approach for the calculation of the $z_g$-distribution here. This is mainly because it is a well-defined theoretical limit which allows for a clear interpretation. Besides, relaxing this assumption in Sec.~\ref{sec:alternative} brings to light strong effects of energy loss that seems to corroborate its relevance.

\section{Jet grooming in the coherence approximation}
\label{sec:Observable}

As alluded to in the introduction, the $z_g$-distribution may provide for the first time a direct measurement of the medium-induced gluon splitting probability. Moreover, it may also be sensitive to the way energy is lost to the medium by the two substructures. 

The range of subjet energies explored in the CMS measurement \cite{CMS:2016jys}, ensures that the primary splitting that created them occurs at very short time scales as compared to the time scale over which energy loss develops, which is typically of order of several fm's. To see this, consider as an example a sample of jets with $p_\text{\tiny T} = 200$ GeV, within the range of measured momentum fractions $z \sim 0.1-0.5$ and jet resolution angles $\Delta R_{12}\sim 0.1 - 0.4$, we thus get formation times within the range 1$0^{-2}$ fm $\lesssim t_\text{f}\equiv 1/(z p_\text{\tiny T}  \Delta R^2_{12})\lesssim $ 1 fm, which is much smaller than the average medium length. 

In this exploratory study we prefer to reduce the complexity of the technical aspects in order to clarify the physics interpretation avoiding the loss of generality. Let us summarise our discussion so far by listing the main assumptions and caveats of our current setup.
\begin{itemize}
\item We consider two radiative mechanisms that are assumed to be well separated in angles. On the one hand, hard and (quasi-)collinear splittings that remain within the jet cone. They can be vacuum or medium-induced splittings, where the latter are assumed to be rare emissions with large formation times, with energies $\omega_s \lesssim \omega \lesssim \omega_c$, and thus can be computed to leading order in the coupling constant. On the other hand, large-angle medium-induced soft radiation, $\omega \lesssim \omega_s$. These gluons are not captured in the jet cone and are responsible for jet energy loss but not for particle number change. This separation presumes therefore that $\omega_s \lesssim z_\cut E$

\item We focus primarily on a scenario where the jets are not resolved by the medium, i.e., $\theta_c \gtrsim R$. The coherent energy loss picture matches closely the experimental ``soft drop declustering''  procedure since the intra-jet structure is not modified. In addition to the energy loss suffered by the jet as a whole we include the emission of medium-induced radiation.

\item We assume that the relevant part of the vacuum cascade occurs at short enough times scales such that one can ignore energy loss prior to the hard splittings. This is, of course, not applicable to medium-induced branchings which can be triggered anywhere along the in-medium jet path.

\item Finally, we focus, for simplicity. on purely gluonic cascades. 

\end{itemize}

\begin{figure}[t]
\centering
\begin{subfigure}[t]{0.5\textwidth}
\centering
\includegraphics[width=\textwidth]{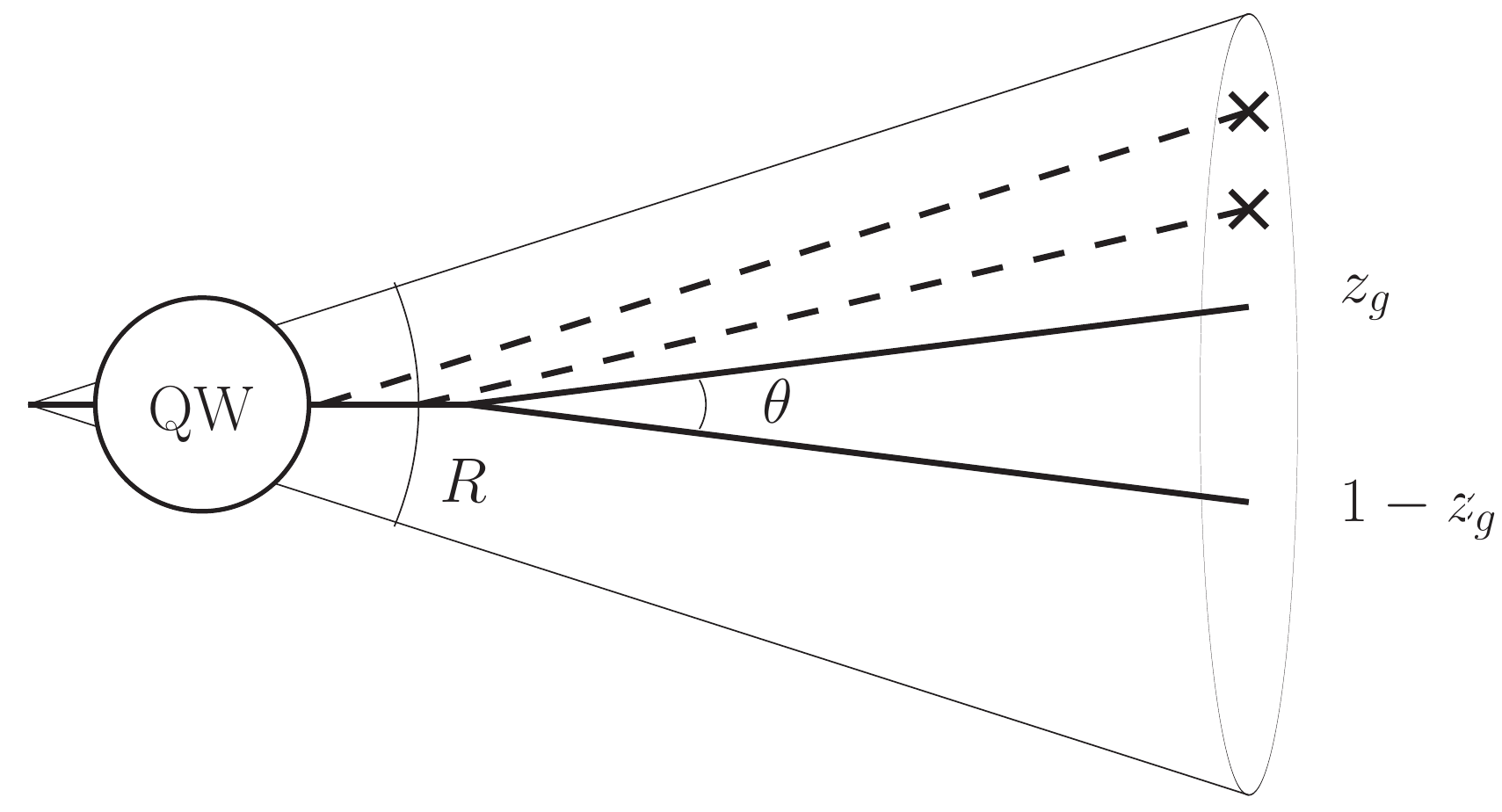}
\caption{}
\label{fig:CohEnLossa}
\end{subfigure}%
~
\begin{subfigure}[t]{0.5\textwidth}
\centering
\includegraphics[width=\textwidth]{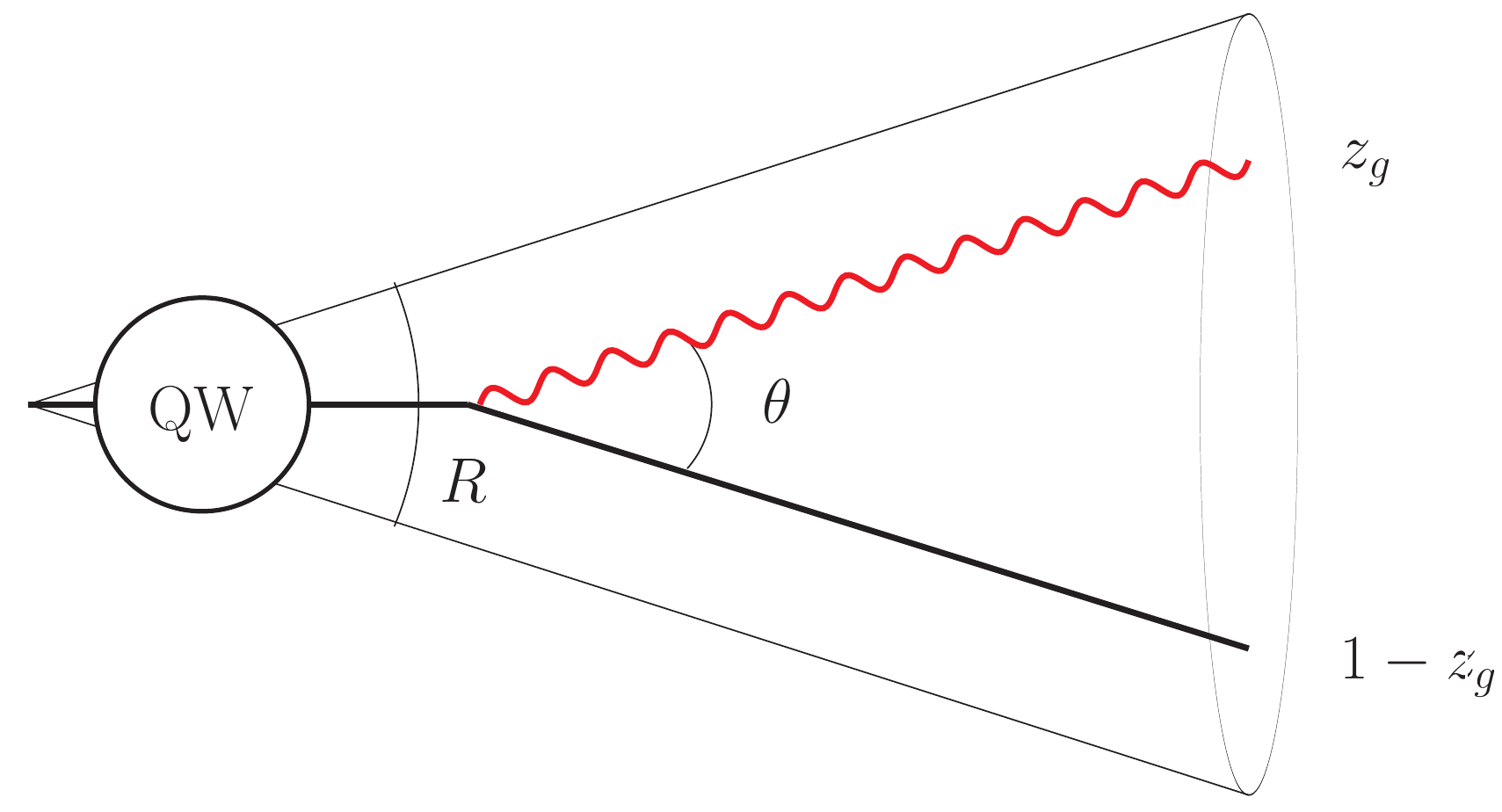}
\caption{}
\label{fig:CohEnLossb}
\end{subfigure}
\caption{Coherent energy loss. The blob on the first propagator illustrates the coherent energy loss of the jet which does not affect the groomed jet structure. Its range extends however along the whole in-medium path length. (a) Vacuum radiation, including real, groomed emissions that are marked by a cross.  (b) Hard, quasi-collinear medium-induced radiation (depicted by a red, wavy line) emitted inside the jet cone.}
\label{fig:CohEnLoss}
\end{figure}

If the jet transverse extension in the medium, $r_\perp \sim RL$, is small compared to the medium resolution scale, $l_\med \sim \sqrt{12/\hat q L}$, it preserves its colour coherence and interacts with the medium as a single charge, see \cite{CasalderreySolana:2012ef} for a discussion. Besides the trivial case, when the jet consists only of one parton, this is explicitly seen in the case when the jet consists of two partons, see e.g. \cite{Casalderrey-Solana:2015bww}. This implies that, even if we include splittings, we can treat the jet as if it only consists of a parent parton carrying the total energy and colour charge of the jet. Hence, both the multiple, soft and rare, hard medium-induced emissions take place off the parent.

For inclusive jets, it is still unclear how to relate exactly the relevant dynamical jet scales to the medium ones. We will therefore assume that the critical angle $\theta_c$ is of the order of the reconstructed radius of the jet, in the sense that the vacuum-like emissions can be treated as colour coherent but the hard emissions still can be emitted within the jet cone. This can be treated as a small departure from the coherence limit.
The two contributions are illustrated in Fig.~\ref{fig:CohEnLoss}. All in-cone emission, including early emissions that fail to pass the grooming condition, marked with an ``$\times$'' in Fig.~\ref{fig:CohEnLossa}, are emitted coherently. 

In this approximation the jet $\pT$ relates to a jet that lost energy. Hence, the number of jets measured is given by the quenching factor in Eq.~(\ref{eq:QuenchingFactor}). It also follows that any collinear splitting would appear as if it occurred after the energy was lost. Therefore, in this case we have
\beq
\label{eq:CoherentSplittingFunction}
\mathcal{P}_\vac^\coh(z,\theta) = \mathcal{P}_\vac(z,\theta) \,,
\eeq
that is the splitting of energy between the two subjets matches the vacuum. This is a manifestation of the fact that coherent energy loss does not resolve nor modify the inter-jet structure \cite{CasalderreySolana:2012ef}. 

In order to write a well defined splitting function, as in Eq.~(\ref{eq:SplitFuncDef}), we have to resum virtual and real emissions that fail to pass the grooming condition into the Sudakov form factor. Since, as described above, currently  all emissions are assumed to be affected coherently by any energy loss, two such emissions are illustrated in Fig.~\ref{fig:CohEnLossa}, the form factor is simply the vacuum one. This procedure clearly defines a consistent way of defining infrared safe observables in the presence of a medium. We explain how the Sudakov form factor would be modified in a more involved situation in \ref{sec:AppendixSudakov}.

The diagram describing hard medium-induced radiation with energies in the range $\omega_s < \omega \lesssim\omega_c$ is given in Fig.~\ref{fig:CohEnLossb} and has been discussed in Sec.~\ref{sec:MediumInduced}. Even though these emissions are rare, $\mathcal{O}(\alpha_s)$, their energies are in the relevant range for the measurement. Taking into account the constraint on a minimal resolution angle, $\Delta R_{12} \geq 0.1$ one should be sensitive to these emissions in jets with $\pT \sim 100-200$ GeV. 
We restore the angular dependence of the spectrum in Eq.~(\ref{eq:BDMPSspectrum}) by taking into account additional Gaussian broadening of the medium-induced particles, described by $\exp[ - \theta^2/ \langle \theta^2 \rangle]/\langle \theta^2 \rangle$, where $\langle \theta^2 \rangle \equiv \hat q  (L-t)/[z_g(1-z_g)\pT]^2$ is the average angular broadening acquired during from the production point $t$ until the gluon leaves the medium \cite{Blaizot:2014ula,Blaizot:2014rla}. Integrating over $t$, from $0$ to $L$, the distribution above and dividing by $L$ we obtain for the broadening probability,
\beq 
P_\text{br}(\theta) = \frac{z^2_g(1-z_g)^2\pT^2}{\hat q L} \Gamma\left(0,\frac{z^2_g(1-z_g)^2\pT^2\theta^2 }{\hat q L} \right) \,,
\eeq
where $\Gamma(0,x)$ is the incomplete gamma function. The characteristic momentum fraction, where this effect plays a role, is however rather small for our purposes. Then, the final formula becomes 
\beq
\label{eq:BDMPSfinal}
\mathcal{P}_\med(z_g,\theta) =\, 2\theta \,P_\text{br}(\theta) \,\frac{\dd N_\BDMPS}{\dd z_g}\, \Theta_\text{cut}(z_g-z_\text{cut}\theta^\beta),
\eeq
where, as for the vacuum splitting, we have assumed that energy loss is coherent and therefore does not affect the observables. For these splittings, the jet spectrum is again modified in the same way as for the vacuum emissions, see above Eq.~(\ref{eq:CoherentSplittingFunction}).

Because of the collinear safety of the BDMPS-Z spectrum no Sudakov form factor is to be associated with it. In physical terms, this means that medium-induced gluons are produced at relatively large angles such that the probability for the de-clustering procedure to count a BDMPS-Z splitting as being the first in the cascade is close to unity. 

Gathering the inputs from the discussion above, the final formula for the splitting probability ($z_g <1/2$) for coherent jets in heavy-ion collisions then reads
\begin{align}
\label{eq:CoherentSplittingTot}
\textsl{p}(z_g) &= \int_{0}^{\thetajet}\dd \theta\, \Delta(R,\theta) \mathcal{P}_\vac(z_g,\theta) \Theta\left(z_g-z_\cut \theta^\beta \right) \left[1-  \int_0^{R} \dd \theta \int_{z_\cut\theta^\beta}^{1/2} \dd z \,\mathcal{P}_\med(z,\theta) \right]\nn
&+ \int_{0}^{\thetajet} \dd \theta  \, \mathcal{P}_\med(z_g,\theta)\Theta\left(z_g-z_\cut \theta^\beta \right)  \,,
\end{align}
where the Sudakov form factor and the two splitting functions can be found in Eqs.~(\ref{eq:Sudakov}), (\ref{eq:CoherentSplittingFunction}) and (\ref{eq:BDMPSfinal}), respectively. 

A few remarks are in order. The first and second terms correspond to the the probability for the two subjets to be formed by a vacuum or a medium-induced splitting, respectively. Now, in order for the measured hard splitting to be vacuum-like, one has to ensure that no rare medium-induced splitting had occurred earlier. This is accounted for by the suppression factor in the square-brackets in the first term that corresponds to the probability of no medium-induced radiation. As a result, the medium-modified splitting function $\textsl{p}(z_g) $, given by Eq.~(\ref{eq:CoherentSplittingTot}), is properly normalised as a probability. 

Again, one of the underlying assumptions leading to the factorised form in Eq.~(\ref{eq:CoherentSplittingTot}) is the angular separation between the vacuum and the medium-induced splittings. The fact that the angular integration for the vacuum part is not sensitive to the upper limit $R$, while the medium-induced contribution is not sensitive to the lower limit, justifies our approximation. Corrections to Eq.~(\ref{eq:CoherentSplittingTot}) are sub-leading in the leading-log approximation. 

\begin{figure}
\centering
\includegraphics[width=0.6\textwidth]{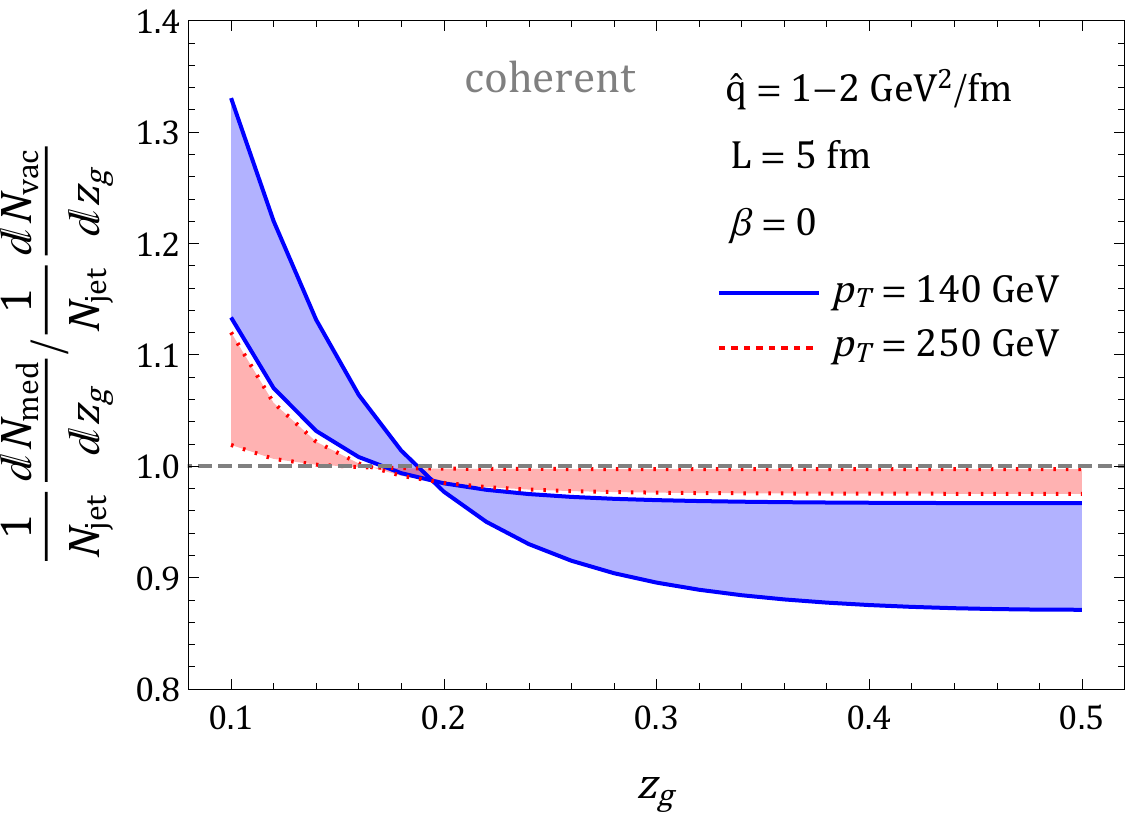}
\caption{(Color online) The ratio of normalised $z_g$-distributions in Pb-Pb and pp collisions for $\pT = 140$ GeV (full lines) and  $\pT = 250$ GeV (dashed lines). The shaded area between the pairs of curves accounts for the variation of $\hat q$.}
\label{fig:CoherentResults}
\end{figure}

The sensitivity to the minimal angle $R_0$ can easily be included in Eq.~(\ref{eq:CoherentSplittingTot}) by replacing $R_0$ in the lower limits of the angular integrals. We have plotted the ratio of the {\it normalised} medium-modified splitting function (\ref{eq:CoherentSplittingTot}) to the vacuum one (\ref{eq:SplitFuncDef}) in Fig.~\ref{fig:CoherentResults} (normalisation to the number of jets). We have considered a static medium of length $L=5$ fm, which is close to the average path-length of jets traversing the medium at LHC, and characterised by a constant transport parameter in the range $\hat q = 1-2$ GeV$^2/$fm, that gauges the uncertainty on the medium parameter, and used $\alpha_s = 0.3$. Finally, we set $R=0.3$ and replace $R_0=0.1$ as in the experimental data. Note that one-pronged jets are discarded in the experimental procedure, hence the distribution in Fig.~\ref{fig:CoherentResults} is self-normalised. 

The two-prong probability (\ref{eq:CoherentSplittingTot}) is a result of the interplay between vacuum radiation that is unaffected by energy loss and BDMPS-Z gluons that are emitted within the cone. Roughly speaking, their $z$-dependence is given by $z^{-1}$ and $z^{-\threehalfs}$, respectively. Since both terms are approximately proportional to the same quenching factor, which scales out of the expression, it is mainly the characteristic energy $\omega_c$ that controls the enhancement. The jet energy dependence of the relative contribution is contained in the BDMPS-Z term, which scales as $\sim\sqrt{\omega_s/E}$ implying a more pronounced enhancement over the pure vacuum spectrum at low jet energies. This is also apparent in Fig.~\ref{fig:CoherentResults}.
The steepening of the distribution for increasing $\hat q$ reflects a larger range for the induced bremsstrahlung, as $\omega_c$ grows, but also the stronger effects of momentum broadening at angles closed to the experimentally minimal resolution. 

Similarly to the vacuum case, introducing a resolution angle $R_0$, the integral of the splitting function  yields the probability to measure two prongs,
\beq
\label{eq:Coh2prong}
\mathbb{P}_{2\text{prong}} = \int_0^{1/2} \dd z_g\, \textsl{p}(z_g) = 1- \left[1-  \int_{R_0}^{R} \dd \theta \int_{z_\cut \theta^\beta}^{1/2} \dd z  \,\mathcal{P}_\med(z,\theta) \right]\, \Delta(R,R_0) \,.
\eeq
We observe that the probability is larger in the medium than in the vacuum because of the presence of a new radiative mechanism, the BDMPS-Z spectrum. Since probability is conserved, $\mathbb{P}_{1\text{prong}}+\mathbb{P}_{2\text{prong}}=1$, $\mathbb{P}_{1\text{prong}}$ is correspondingly smaller.
\begin{figure}
\centering
\includegraphics[width=0.6\textwidth]{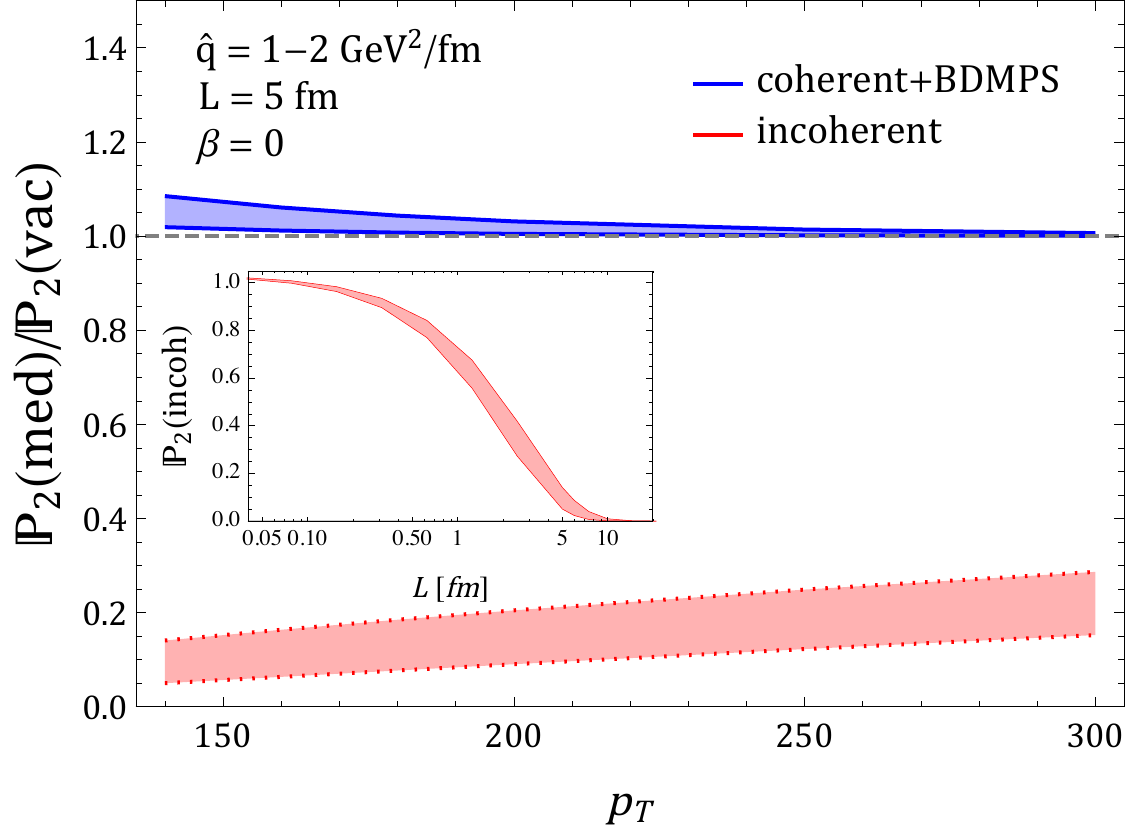}
\caption{(Color online) The ratio of the two-pronged probability for two scenarios described in the text compared to the vauum. In the inset, we plot the two-pronged probability for incoherent energy loss at for the lowest $\pT$-bin as a function of varying the medium length.}
\label{fig:ResultsTwoPronged}
\end{figure}
We plot the ratio of the two-pronged probability for coherent jets (\ref{eq:Coh2prong}) to the corresponding vacuum probability (\ref{eq:2prongVacuum}) in Fig.~\ref{fig:ResultsTwoPronged} for a the same set of parameters as in Fig.~\ref{fig:CoherentResults} (blue curves). We observe a modest enhancement of the two-pronged probability in heavy-ion collisions at small transverse momenta that is completely driven by rare BDMPS-Z emissions. The smallness of the effect warrants a perturbative treatment of this component. We have also checked that the result is not sensitive to the choice of $R_0$, due to the preferred large-angle radiation of the BDMPS-Z spectrum. In the same figure, we also plot the same ratio of jet affected by incoherent energy-loss, see Sec.~\ref{sec:incoherent} and Eq.~(\ref{eq:Incoh2prong}), that will be discussed shortly.

\section{Subtleties of jet grooming in heavy-ion collisions}
\label{sec:alternative}

Above, we have given a detailed account of a situation where the jet remains coherent in the medium. It sources induced bremsstrahlung via its total colour charge, part of which remains within the cone. In order to substantiate the main message of the paper, namely the interplay of jet coherence and hard radiation, let us scrutinise the possibility of disregarding both. Firstly, let us neglect effects of coherence so that all constitutes of the jet interact with the medium independently. Secondly, let us assume that there is no perturbative component of the induced radiation that can be emitted at small angles. We are then left with the possibility that the jet can degrade its energy through soft emissions (or, equivalently, interactions through drag). In parallel, back-reaction from the medium can source a wake that drives recoil partons from the medium into the reconstructed jet cone. In effect, this is equivalent to a net energy {\it gain}. These considerations shed new light on the sensitivity of the $z_g$-distribution to soft medium properties, and adds to the importance of our main physics message, presented in Sec.~\ref{sec:Observable}.

\subsection{Independent energy loss}
\label{sec:incoherent}

Let us for the moment assume that all {\it primary} subjets in the cone lose energy independently. We consider that all vacuum emissions from the jet parent interact independently with the medium. Still, each of these substructures are separately unresolved by the medium. This simplification is sufficient to capture the crucial, qualitative differences between the coherent scenario discussed above and the incoherent one. Also, while we explicitly calculate the process using quenching weights, the physics result is completely generic and applies to any mechanism leading to energy degradation along the jet path.

The quenching applies solely to the daughters of the splitting process since they are created early in the medium such that the quenching of the parent can be neglected, see Sec.~\ref{sec:Observable}. Nevertheless, a major complication arises in this scenario. While the jet yield factorises from the jet evolution in the coherent and vacuum cases, in this case it does not because the final jet $\pT$ does not match the energy of the jet at its formation point. As a result, the observable will depend on the jet spectrum.  

Let us first consider the situation where the resolution angle $R_0$ is not small compared to the jet opening angle and hence, the angular integration does not yield a large logarithm that should be resummed otherwise. In this case,  it is sufficient to restrict our discussion to the leading order which amounts to neglecting the Sudakov. Using jet calculus we find that the two-gluon probability reads
\begin{align}
\label{eq:IncoherentVacuumDef1}
\frac{\dd N ( \pT)}{\dd  \pT^2}   \textsl{p}(z_g) &=\,\abar \ln\frac{R}{R_0}\int_0^\infty \dd \epsilon \int_0^\epsilon \dd \epsilon' \, \,\Theta(z_g-z_\cut\theta^\beta)\nonumber \\
&\times  D_\QW(\epsilon-\epsilon')D_\QW(\epsilon')\frac{ \pT}{ \pT+\epsilon} P\left(\frac{z_g \pT+ \epsilon'}{ \pT+\epsilon}\right) \frac{\dd N ( \pT+\epsilon)}{\dd  \pT^2} \,,
\end{align}
for $z_g <1/2$.
Let us presently point to two main features of this expression. We can expand the Altarelli-Parisi splitting function for $\epsilon$ and $\epsilon' \ll \pT$, keeping only the second order in $\epsilon'$, that is sensitive to the small-$z_g$ region as follows,
\begin{align}
P\left(\frac{z_g \pT+ \epsilon'}{\pT+\epsilon}\right) & \simeq P\left(z_g\right)  + P'\left(z_g\right) \frac{\epsilon'}{\pT}\simeq \frac{1}{z_g} \left(1-\frac{\epsilon'}{z_g \pT}\right) \,.
\end{align}
Since $\epsilon'$ typically is determined by the medium scale $\omega_s$, it follows that for large $z_g$ the two-prong splitting function tends to the vacuum one, while at small $z_g$ it is suppressed. Hence, we observe that the $z$-spectrum experiences a characteristic shift $z \to z + z^\loss$, where the characteristic shift can only be a ratio of the relevant scales of the problem $z^\loss \sim \omega_s/\pT$. This results in a flattering of the $z_g$-distribution in contrast to what is observed in the data \cite{CMS:2016jys}. We emphasise that this is a generic feature which does not depend on the specific form of the basic splitting function nor on the mechanism of energy degradation after the splitting takes place. The second difference concerns the quenching weights that, in contrast to the coherent case, appears twice. 
Taking as an extreme case the situation where each quenching weight gives rise to a corresponding quenching factor (for $n\gg 1$), this result implies that, supposing that all jets consist of two resolved structures, the relevant medium parameters should be reduced by a factor $\sim 4$ in order to obtain the same quenching of the inclusive jet spectrum, see Eq.~(\ref{eq:QW-prob}). In fact, as the discussion below will disclose, for asymmetrical splittings the dominant effect turns out to be the quenching of the jet ``core'', which in our case corresponds simply to the leg with the largest transverse momentum, will dominate the overall quenching of the two-pronged system.

Let us now turn to the situation where $\abar \ln R/R_0 \gg 1$. In this case one has to resum multiple vacuum emissions into the Sudakov form factor.  A detailed discussion of this case is postponed to future work. However, one can derive a closed formula in the  limit where the energy that is groomed away is much smaller than the jet energy, i.e., $\pTg \simeq  \pT$ and for small energy loss, $\pT \gg \epsilon$, such that one can neglect $\epsilon$ everywhere except in the steeply falling spectrum in Eq.~(\ref{eq:IncoherentVacuumDef1}). 
Furthermore, we shall assume that the quenching of the groomed energy is small compared to the quenching of the total jet energy, see \ref{sec:AppendixSudakov} for further details. For the moment we do not consider additional BDMPS-Z emissions. With these assumptions, the gluon splitting probability reads
\begin{align}
\label{eq:IncoherentVacuumDef2}
\frac{\dd N_\jet}{\dd \pT^2}  \,\textsl{p}(z_g) &=  \int_0^{R} \dd \theta\int_0^\infty \dd \epsilon \int_0^{(1/2-z_g)\pT} \dd \epsilon'  \,\Delta(R,\theta|\pT) \Theta(z_g-z_\cut\theta^\beta) \nonumber \\
&\times  D_\QW(\epsilon)D_\QW(\epsilon') \mathcal{P}_\vac\left(z_g+\frac{\epsilon'}{ \pT},\theta \right) \,\frac{\dd N_{\jet(0)}( \pT + \epsilon)}{\dd \pT^2}\,,
\end{align}
for $z_g < 1/2$ and $z_g+\epsilon_2/\pT <1/2$ since we assumed asymmetric splittings prior to energy loss.
The medium modified Sudakov form factor now also depends on the final jet energy and reads
\beq
\label{eq:IncoherentSudakovDef}
\Delta(R,R_0|\pT) = \exp\left[  - \int_{R_0}^R \dd \theta\int_{z_\cut\theta^\beta}^{1/2} \dd z  \int_0^{(1/2-z)\pT} \dd \epsilon \,D_\QW(\epsilon)\,{\cal P}_\vac\left(z+\frac{\epsilon}{\pT},\theta \right) \right] \,.
\eeq
It is staightforward to verify that in the limit $\pT\to \infty$ we recover the formula for the Sudakov in the vacuum (\ref{eq:SplitFuncDef}).

Introducing again a minimal resolution angle $R_0$, the probability of observing a two-pronged structure now reads
\begin{align}
\label{eq:Incoh2prong}
\mathbb{P}_{2\text{prong}} &\equiv  \int_0^{1/2} \dd z_g \, \textsl{p}(z_g) =1-\Delta(R,R_0|\pT) \,,
\end{align}
where we have taken advantage of Eq.~(\ref{eq:NumberofQuenchedJets}). In contrast to the scenario described in the previous section, since we did not consider an additional mechanism that could emit inside the cone, but only included energy loss, we expect $\mathbb{P}_{2\text{prong}}$ to be smaller than for Eq.~(\ref{eq:Coh2prong}). Indeed, the exponent of the medium modified Sudakov (\ref{eq:IncoherentSudakovDef})
is suppressed by a quenching weight causing the two-pronged probability to decrease. For instance, in the extreme case scenario where $\epsilon/\pT > 1- z_\cut\theta^\beta$, all sub-jets fall below the cut and $\Delta(R,R_0|\pT)=1$, resulting in the vanishing of the two-pronged probability.

The dramatic effect described above is clearly seen in Fig.~\ref{fig:ResultsTwoPronged} (red curves). In the main figure we plot the ratio of the incoherent two-pronged probability (\ref{eq:Incoh2prong}) to the vacuum probability (\ref{eq:2prongVacuum}) for the same parameters as used in Fig.~\ref{fig:CoherentResults}. We observe a large suppression, of the order of a factor $\sim5 - 10$, that slowly vanishes with increasing $\pT$. In the inset we probe the sensitivity of (\ref{eq:Incoh2prong}) to varying medium parameters, in this case the medium length $L$, and point to a strong sensitivity. We conclude that this observable clearly sets apart the expectation from the coherent and incoherent jet scenarios. 

We emphasise that the features we have discussed are generic and not dependent on the specific scenario (quenching weights). They should therefore be present in any model that applies incoherently energy loss to all constituents of a jet, e.g., in a Monte-Carlo energy loss model based on the vacuum shower \cite{Casalderrey-Solana:2014bpa}.

\subsection{Sensitivity to soft particles}
\label{sec:recoil}

In direct analogy with the medium-induced component that remains within the jet cone, one could also consider a possible mechanism of jet energy {\it gain}. In contrast to the former, it is currently a subject of theoretical debate and various implementations exist, see, e.g.  \cite{Zapp:2012ak,Wang:2013cia,He:2015pra,Casalderrey-Solana:2016jvj}. 

The soft drop does not filter out all soft particles from the jet, merely the ones that are well separated from the jet core. That implies that the procedure could be sensitive to the addition of random soft particles that are located close in angle to the two jet substructures that are singled out. We can, for instance, imagine this effect to be caused by medium back-reaction to the jet propagation. In this case, in contrast to the energy loss effect, a net energy {\it gain} would lead to the opposite shift $z \to z - z^\gain$ in the splitting function, cf. Eq.~(\ref{eq:IncoherentVacuumDef1}). As in the previous case, $z^\gain$ can be estimated to be ratio of the energy of residual background fluctuations and the jet energy. Because of the rapid rate of rescattering of medium partons that cause their diffusion to large angles, we expect $z^\gain$ to be sensitive only to fluctuations that have been generated close to the surface of the medium and therefore the residual effect should not be enhanced by the size of the medium, thus $z^\gain \ll z^\loss$. Furthermore, we expect the energy dependence to be much slower than the one in Eq.~(\ref{eq:CoherentSplittingTot}) since there is only one component. Monte Carlo event generators that include medium recoil effects, such as \cite{Zapp:2012ak,Wang:2013cia,He:2015pra,Casalderrey-Solana:2016jvj}, should be sensitive to this modification. 

It is worth noticing that the sensitivity to the soft gluons in the jet cone was already noticed for jets in proton-proton collisions \cite{Dasgupta:2015yua}, and should therefore be crucial at low-$p_\text{\tiny T}$. Part of the effect is therefore cancelled when taking the ratio of heavy-ion data to proton-proton, as in Fig.~\ref{fig:CoherentResults}, see also \cite{CMS:2016jys}. One can also enhance the grooming procedure to suppress this contribution even more \cite{Dasgupta:2015yua}. Nevertheless, this sensitivity calls to interpret the results of jet substructure measurements with due care.

\section{Summary and outlook}
\label{sec:outlook}

The study of jet modifications in heavy-ion collisions is still in its early stages, and comprises, e.g., papers on jet shapes \cite{Salgado:2003rv,Polosa:2006hb,Vitev:2008rz,Chien:2014nsa,Chien:2015hda} and fragmentation functions \cite{Armesto:2007dt,Mehtar-Tani:2014yea}. In the current work, we consider the effects of the ``soft drop declustering'' on jets in medium and their resulting distribution of the energy-sharing fraction variable $z_g$. Uniquely so far in the context of jet observables in heavy-ion collisions, this procedure calls for a treatment of jet substructures, their formation and their evolution alongside medium effects. Moreover, energy loss in the medium gives the jet grooming a physical meaning. In particular, one easily realises that emissions which a priori would pass the grooming criterium in vacuum could fail to do so in the medium due to energy loss. Besides, medium-induced radiation, provided it is sufficiently hard, could potentially be identified as new structures on the subjet level.

In this work, we have, for the first time, considered the effect of energy loss on jet substructures and shed light on the important interplay of medium-radiation outside and within the jet cone. 

Let us summarise by recapping the main novelties of our work: 
\begin{enumerate}
\item We have argued that a plausible explanation to the observed features in \cite{CMS:2016jys} are rare, hard BDMPS-Z emissions off mainly coherent jets. Ignoring BDMPS-Z emissions or color coherence, we obtain either weaker or completely opposite trends. 

We should, however, get a better handle on the sensitivity to soft gluons at the medium scale, see Sec.~\ref{sec:recoil}.
While the mechanism of energy loss is a crucial ingredient for describing a well of inclusive observables, jet substructure measurements enjoy a unique sensitivity to the direct emission of medium-induced gluons.
\item  We have calculated consistently the probability of having a two- or one-pronged jets that is yet to be measured experimentally (see Fig.~\ref{fig:ResultsTwoPronged}, and Eq.~(\ref{eq:Coh2prong}) and discussion below) and noted that, in the presence of additional BDMPS-Z radiation, the former grows for coherent jets, as expected intuitively. In contrast, in the case where sub-jets lose energy independently to large angles, the two-pronged probability is strongly suppressed. Therefore, we expect this observable to be sensitive to various mechanisms of in-medium jet interactions. 

\end{enumerate}

A more careful treatment of coherence effect in the jet fragmentation should naturally involve an interpolation in the grooming angle between the two extreme scenarios described in Secs.~\ref{sec:Observable} and \ref{sec:incoherent}.
Developing these methods further, aiming for a quantitative analysis, would also allow to validate various probabilistic Monte Carlo prescriptions that are currently used in the analysis of heavy-ion data.
We leave these very interesting aspects for an upcoming publication.

\section*{Acknowledgments}
We acknowledge discussions with Marta Verweij and Phil Harris about the experimental aspects of the measurement reported in \cite{CMS:2016jys} and with Guilherme Milhano, Urs Wiedemann and Jorge Casalderrey-Solana.
KT has been supported by a Marie Sk\l{}odowska-Curie Individual Fellowship of the European Commission's Horizon 2020 Programme under contract number 655279 ResolvedJetsHIC. The research of YMT is supported by the U.S. Department of Energy under Contract No. DE-FG02-00ER41132.

\appendix
\section{The Sudakov form factor}
\label{sec:AppendixSudakov}
Considering the procedure of jet grooming via soft drop, going from the reconstructed jet cone to a minimal resolution angle, we realise that the outcome can only be one of two: either we find a splitting or not, which implies that it took place at too small angles to be resolved. These outcomes are, simply, the possibility of finding a two-pronged or a one-pronged jet. Let us therefore take advantage of this fact to say something about the total probability.
Hence, given a minimal resolution angle $R_0$, the total probability of either observing a two- or one-pronged jet, $\textsl{p}_\tot(z_g) = \textsl{p}^{(2)}(z_g) + \textsl{p}^{(1)}(z_g)$, reads
\beq
\label{eq:TotalProbabilityDef}
\textsl{p}_\tot(z_g) =   \frac{1}{2}\int_{R_0}^R \dd \theta\, \Delta(R,\theta) {\cal P}_\vac(z_g,\theta)\Theta_\text{cut}(z_g,\theta)  + \delta(1-z_g) \Delta(R,R_0),
\eeq
in vacuum, where $\Theta_\cut(z,\theta) \equiv \Theta(1/2-z)\Theta(z-z_\cut\theta^\beta) + \Theta(z-1/2)\Theta(1-z-z_\cut \theta^\beta)$. Note that this expression does not depend on the initial spectrum of jets. Taking advantage of the symmetrisation $ \mathcal{\bar P}_\vac(z,\theta) \equiv  \mathcal{P}_\vac(z,\theta) +  \mathcal{P}_\vac(1-z,\theta) =  2\mathcal{P}_\vac(z,\theta)$ and limiting the range of groomed momentum fractions to $z_g <1/2$ automatically decouples the last term and we recover the expression in Eq.~(\ref{eq:SplitFuncDef}).

In continuation, the last term in Eq.~(\ref{eq:TotalProbabilityDef}) allows to determine the proper Sudakov form factor for the process by the following procedure. The integral of the $z_g$-probability distribution must be normalised and independent of $R_0$. Hence, 
\beq
\label{eq:ProbConservation1}
\frac{\dd}{\dd R_0} \, \int_0^1 \dd z_g\, p(z_g) =0 \,,
\eeq
from which it follows that 
\beq
\label{eq:evol-sudakov-vac}
\frac{\dd}{\dd R_0} \Delta(R,R_0)=  \frac{1}{2}\int_0^1 \dd z \,{\cal P}_\vac(z,R_0)\Theta_\cut(z,R_0)  \Delta(R,R_0) \,. 
\eeq
It is straightforward to solve this equation and one finds 
\beq
 \Delta(R,\theta)=  \exp\left[ - \frac{1}{2} \int_\theta^R \dd \theta' \int_0^1 \dd z\, {\cal P}(z,\theta')\Theta_\text{cut}(z,\theta')\right] \,,
\eeq
as expected. Since the splitting function for coherent jets does not depend on the initial jet spectrum as well, the same procedure can be used to find the conventional Sudakov form factor related to vacuum radiation. However, as discussed in Sec.~\ref{sec:Observable}, the BDMPS-Z spectrum would lead to an additional suppression.

Let us now turn to determining the modification of the Sudakov form factors in the presence of incoherent jet quenching effects, see Sec.~\ref{sec:incoherent} for a further discussion. Following the method above, the total probability, which now depends on the jet spectrum, is again a sum of the probabilities to find one- and two-pronged structures inside the jet. We point out that they are modified in a different way, since the medium resolves every substructure available in the two cases (here we assume that the medium cannot further resolve any structure within the resolution angle, for a discussion see Sec.~\ref{sec:Observable}).

\begin{figure}
\centering
\includegraphics[width=0.55\textwidth]{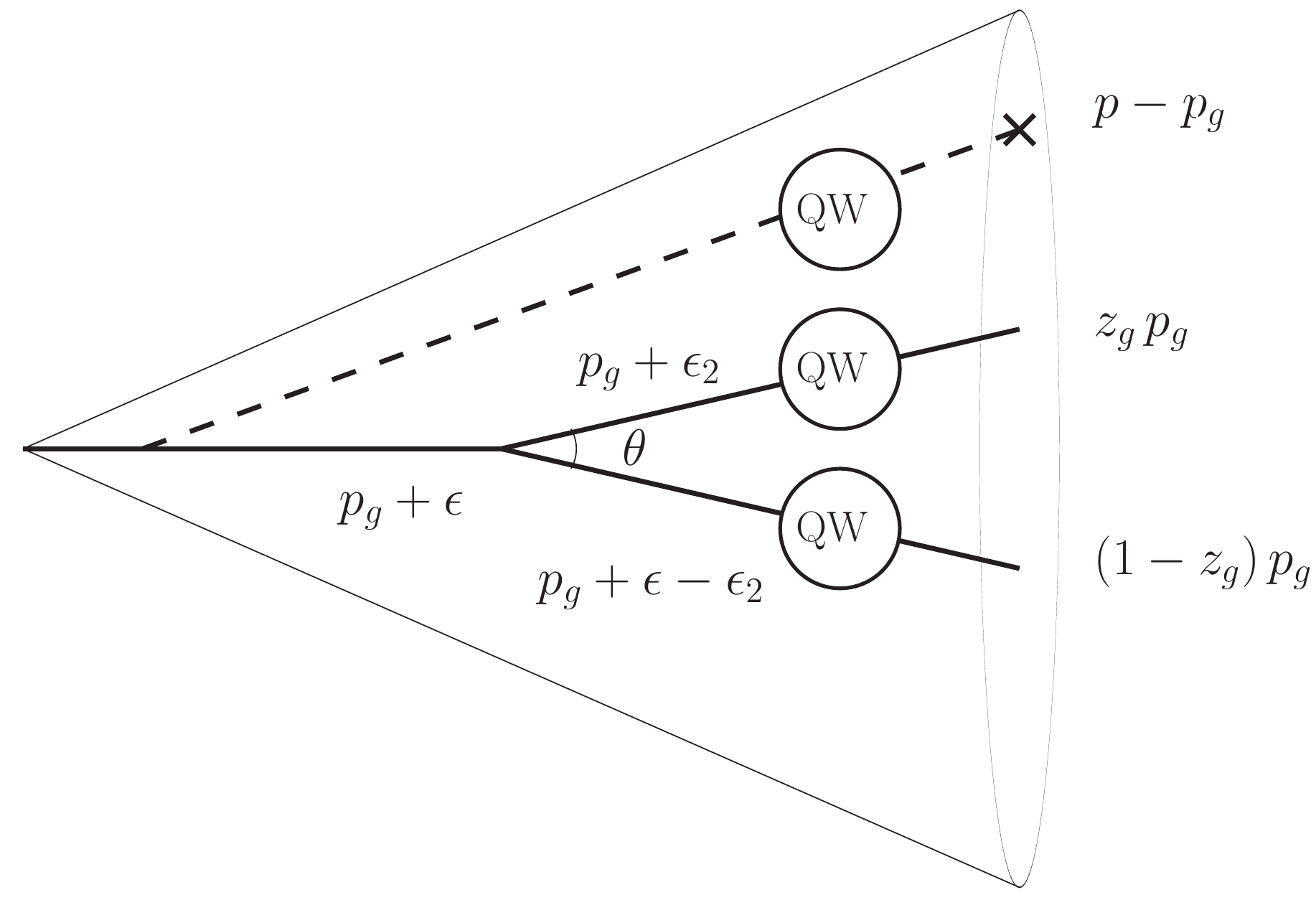}
\caption{Illustration of the splitting function in the incoherent approximation.}
\label{fig:inCohEnLoss}
\end{figure}

This turns out to be significantly more complicated because of the sensitivity to the jet spectrum. The corresponding Sudakov form factor will depend on the energy as well, see also Fig.~\ref{fig:inCohEnLoss}.
Proceeding as before, we find that the probability of finding a one-pronged jet reads,
\begin{align}
\frac{\dd N_\jet}{\dd \pT^2} \textsl{p}^{(1)}(z_g) &= \delta(1-z_g)\,\int_0^{\pT^2}\dd p^2_{g}\, \int_0^\infty\dd \epsilon\, \Delta(R,R_0|\pT - \pTg, \pTg+\epsilon)  \,D_\QW(\epsilon) \,,
\end{align}
where $\pT-\pTg$ stands for the energy that is groomed away, $\pTg+\epsilon$ is the energy of the hard parent before losing $\epsilon$  amount of energy. The groomed jet energy is limited by the total available jet energy.
Here we have introduced a generalised Sudakov distribution, whose initial condition (at $R=R_0$) reads
\beq
\label{eq:initial-Sudakov-indep}
\Delta(R,R|\pT-\pTg,\pT+\epsilon)=\delta(\pT^2-\pTg^2)\,\frac{\dd N_{\jet(0)}(\pT + \epsilon)}{\dd \pT^2}.
\eeq
The expression for the two-pronged probability can be found using our extension of jet calculus \cite{Konishi:1979cb} (see Figure~\ref{fig:inCohEnLoss} for an illustration) and reads
\begin{align}
\label{eq:2-prong-indep}
\frac{\dd N_\jet}{\dd \pT^2}  \,\textsl{p}^{(2)}(z_g) &=  \int_0^{\pT^2}\dd \pTg^2 \int_{R_0}^{R} \dd \theta\int_0^\infty \dd \epsilon \int_0^\epsilon \dd \epsilon' \, \,\Delta(R,\theta | \pT-\pTg, \pTg+\epsilon) \Theta(z_g-z_\cut\theta^\beta)\nonumber \\
&\times  D_\QW(\epsilon-\epsilon')D_\QW(\epsilon')\frac{\pTg}{\pTg+\epsilon} \mathcal{P}_\vac\left(\frac{z_g \pTg+ \epsilon'}{\pTg+\epsilon},\theta \right) \,
\end{align}
for $z_g<1/2$, where $\epsilon'$ and $\epsilon-\epsilon'$ are the energies lost by the $z_g$ and $1-z_g$ prongs, respectively. 

Deriving an evolution equation for the generalised Sudakov form factor is beyond the scope of this work. However, some insight can be gained on the behaviour to expect from Eq.~(\ref{eq:2-prong-indep}) by assuming that: i)  the jet spectrum is a steeply falling function of $\pT$. This amounts to neglecting $\epsilon$ (that is $\epsilon \ll \pTg$, but $ \epsilon_2 \lesssim z_g \pTg$) everywhere except in the Sudakov, which involves the jet spectrum; ii) for asymmetric splittings ($z_g \ll 1$ or $1-z_g\ll 1$) the total quenching should be dominated by the ``hardest'' leg implying that $\epsilon_2 < \epsilon$; iii) the energy groomed away is small compared to the jet energy, that is,  $p\simeq p_g$.

As a result of the first and second approximations, the integral over $\epsilon$ factorises as follows
\begin{align}
\label{eq:2-prong-indep-approx}
\frac{\dd N_\jet}{\dd \pT^2}  \,\textsl{p}^{(2)}(z_g) 
&= \int_0^{\pT^2}\dd \pTg^2\,  \int_{R_0}^{R} \dd \theta   \int_0^\infty \dd \epsilon \,D_\QW(\epsilon)\Delta(R,\theta | \pT-\pTg, \pTg+\epsilon) \Theta(z_g-z_\cut\theta^\beta)\nonumber \\
&\times  \int_0^{(1/2-z)\pTg} \dd \epsilon' \, D_\QW(\epsilon') \mathcal{P}_\vac\left(z_g+\frac{\epsilon'}{\pTg},\theta \right) \,,
\end{align}
for $z_g>1/2$.
The third approximation allows one to express the Sudakov in the following form,
\beq
\Delta(R,\theta|\pT-\pTg, \pTg+\epsilon) \simeq \delta\left(\pT^2-\pTg^2\right) \Delta(R,\theta|\pT) \, \frac{\dd N_{\jet(0)}(\pT+\epsilon)}{\dd \pT^2} \,.
\eeq
For the total probability, we finally obtain 
\begin{align}
\label{eq:2-prong-indep-approx-2}
&\frac{\dd N_\jet}{\dd \pT^2}  \,\big[ \textsl{p}^{(1)}(z_g)+ \textsl{p}^{(2)}(z_g) \big] = \int_0^\infty \dd \epsilon \,D_\QW(\epsilon)\frac{\dd N_{\jet(0)}(\pT+\epsilon)}{\dd \pT^2}  \Big[\delta(1-z_g) \Delta(R,R_0| \pT)\nonumber \\
&\left. +\Theta\left(\frac{1}{2} - z_g\right) \int_{R_0}^{R} \dd \theta   \int_0^{(1/2-z)\pT} \dd \epsilon' \,D_\QW(\epsilon')\mathcal{P}_\vac\left(z_g+\frac{\epsilon'}{\pT},\theta \right) \Delta(R,\theta|\pT) \Theta(z_g - z_\cut\theta^\beta) \right] \,.
\end{align}
It can be checked that the quantity between brackets is normalised to 1, which allows one to identify,
\beq
\label{eq:NumberofQuenchedJets}
\frac{\dd N_\jet}{\dd \pT^2}  = \int_0^\infty \dd \epsilon D_\QW(\epsilon) \frac{\dd N_{\jet(0)}(\pT+\epsilon)}{\dd \pT^2} \,. 
\eeq
Hence, the sum of the single- and two-prong probability distributions reads
\begin{align}
&\textsl{p}(z_g) \equiv \textsl{p}^{(1)}(z_g)+ \textsl{p}^{(2)}(z_g) \simeq \delta(1-z_g) \Delta(R,R_0|\pT)\nonumber \\
&+\Theta\left(\frac{1}{2}-z_g \right) \int_{R_0}^{R} \dd \theta   \int_0^{(1/2-z)\pT} \dd \epsilon' \,D_\QW(\epsilon')\mathcal{P}_\vac\left(z_g+\frac{\epsilon'}{\pT},\theta \right) \,\Delta(R,\theta|\pT) \Theta(z_g-z_\cut\theta^\beta) \,.
\end{align}
Following the same procedure as in Eq.~(\ref{eq:ProbConservation1}) and below, we finally obtain the Sudakov form factor in Eq.~(\ref{eq:IncoherentSudakovDef}).

\end{document}